# Real-World Evaluation of Protocol-Compliant Denial-of-Service Attacks on C-V2X-based Forward Collision Warning Systems

**Jean Michel Tine***
Ph.D. Student
Glenn Department of Civil Engineering
Clemson University, Clemson, SC 29634, USA
Email: jtine@clemson.edu

**Mohammed Aldeen**
Ph.D. Student
School of Computing
Clemson University, Clemson, SC 29634, USA
Email: mshujaa@clemson.edu

**Abyad Enan**
Ph.D. Student
Glenn Department of Civil Engineering
Clemson University, Clemson, SC 29634, USA
Email: aenan@clemson.edu

**M Sabbir Salek, Ph.D.**
Senior Engineer
USDOT National Center for Transportation Cybersecurity and Resiliency (TraCR)
414A One Research Dr, Greenville, SC 29607, USA
Email: msalek@clemson.edu

**Long Cheng, Ph.D.**
Associate Professor
School of Computing
Clemson University, Clemson, SC 29634, USA
Email: lcheng2@clemson.edu

**Mashrur Chowdhury, Ph.D., P.E.**
Eugene Douglas Mays Chair of Transportation
Glenn Department of Civil Engineering
Clemson University, Clemson, South Carolina, 29634
Email: mac@clemson.edu

*Corresponding author

## ABSTRACT

Cellular Vehicle-to-Everything (C-V2X) technology enables low-latency, reliable communications essential for safety applications such as a Forward Collision Warning (FCW) system. C-V2X deployments operate under strict protocol compliance with the 3rd Generation Partnership Project (3GPP) and the Society of Automotive Engineers Standard (SAE) J2735 specifications to ensure interoperability. This paper presents a real-world testbed evaluation of protocol-compliant Denial-of-Service (DoS) attacks using User Datagram Protocol (UDP) flooding and oversized Basic Safety Message (BSM) attacks that exploit transport and application-layer vulnerabilities in C-V2X. The attacks presented in this study use valid, standard-compliant transport and application-layer traffic to stress the receiver while preserving normal PC5 safety communication. In the implemented testbed, the dominant degradation was observed at the receiver host/application path, where high-rate UDP/IP and oversized valid messages increased queue buildup, CPU contention, and FCW processing delay without requiring malformed traffic or direct saturation of the PC5 radio interface, fully adhering to 3GPP and SAE J2735 specifications, but at abnormally high rates and with oversized payloads that overload the receiver resources without breaching any protocol rules such as IEEE 1609. Using a real-world connected vehicle testbed with commercially available On-Board Units (OBUs), we demonstrate that high-rate UDP flooding and oversized payload of BSM flooding can severely degrade FCW performance. Results show that UDP flooding alone reduces packet delivery ratio by up to 87% and increases latency to over 400ms, while oversized BSM floods overload receiver processing resources, delaying or completely suppressing FCW alerts. When UDP and BSM attacks are executed simultaneously, they cause near-total communication failure, preventing FCW warnings entirely. These findings reveal that protocol-compliant communications do not necessarily guarantee safe or reliable operation of C-V2X-based safety applications. These findings highlight a critical security gap in current C-V2X implementations and underscore the need for future research into robust defense mechanisms to safeguard safety applications against stealthy, standards-compliant flooding attacks.

## INTRODUCTION

Modern transportation systems are undergoing a significant shift as vehicles evolve from isolated actors into interconnected agents capable of communicating with other vehicles, road infrastructure and vulnerable road users. This connectivity is realized through Cellular Vehicle-to-Everything (C-V2X), a transformative wireless technology that supports Vehicle-to-Vehicle (V2V), vehicle to Infrastructure (V2I), Vehicle to Pedestrian (V2P) and Vehicle to Network (V2N) communications (*1*). The specifications for the C-V2X communications begin with the release 14 of the 3rd Generation Partnership Project (3GPP) (*2*). This release comprises two new modes of Long-Term Evolution (LTE) operation (i.e., Mode 3 and Mode 4) which differ from their resource allocation. As shown in Figure 1, Mode 3 relies on the cellular base-station to perform resource allocation. On the other hand, Mode 4 enables vehicles to allocate resources on their own, using the PC5 sidelink interface, which is a cellular technology defined in 3GPP standards that enables direct communication with other nodes (e.g., connected vehicles, Roadside Unit (RSU)). Through this PC5 sidelink, connected vehicles broadcast Basic Safety Message (BSMs), typically at 10Hz, containing critical real time data, such as position, speed, heading, and brake status. These messages are designed in agreement with the Society of Automotive Engineers (SAE) J2735 standard (*3*), hence forming the foundation for real-time safety applications, like Forward Collision Warning (FCW), Emergency Electronic Brake Light (EEBL) alerts,

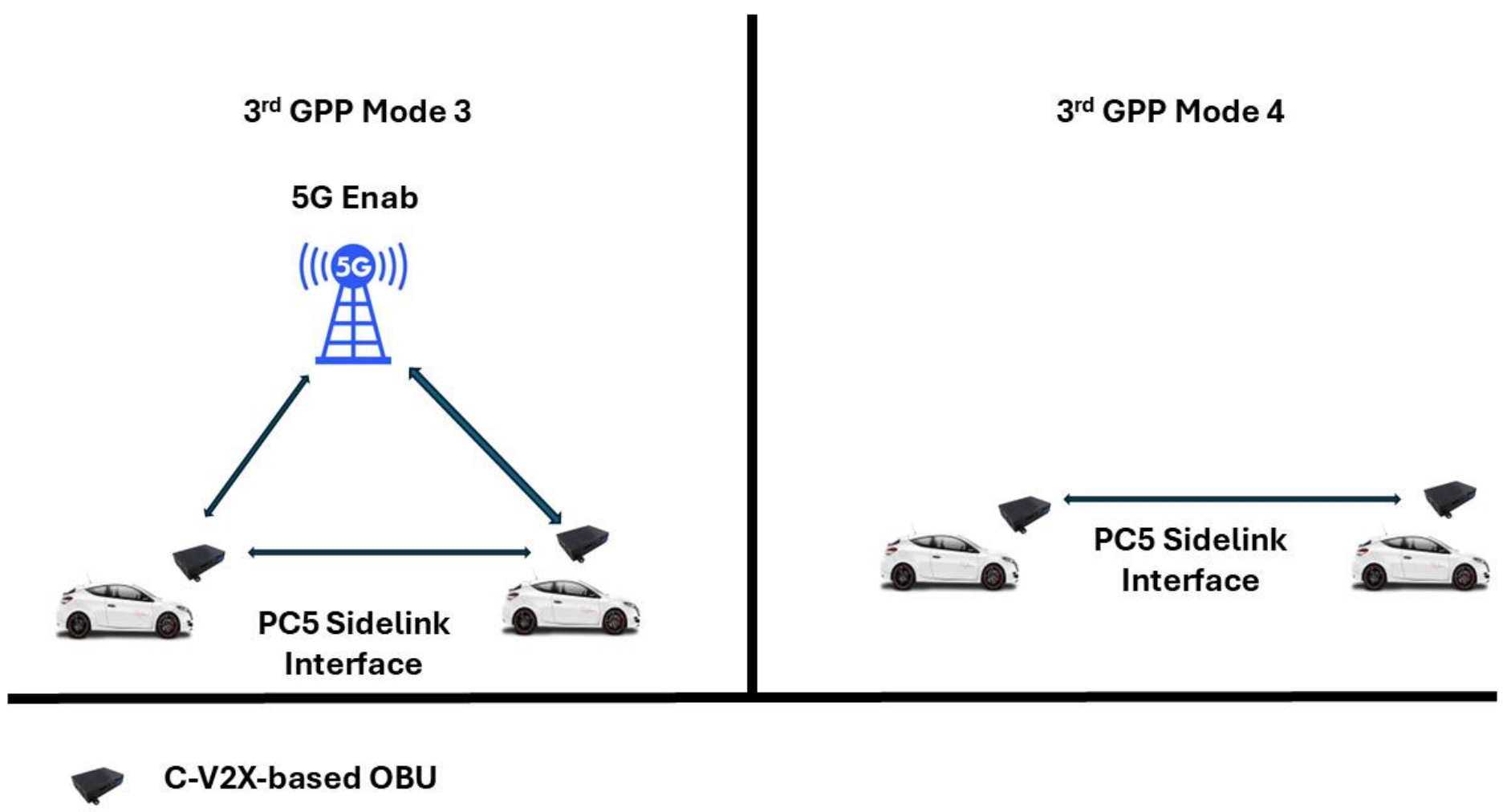

**Figure 1. Modes 3 and 4 of C-V2X Communication**

and intersection movement assist.

The design goal of C-V2X focuses on low latency (under 50 ms) and high packet delivery ratio (PDR) of at least 90% to meet the timing requirements of connected vehicle safety applications (*4*),( *5*). These features make it a more capable successor to the Dedicated Short Range Communication (DSRC) technology, following IEEE 802.11p, by offering greater range and improved reliability under different channel conditions (*4*), (*6*). However, as opposed to Mode 3, where a central base station arbitrates resource allocation, C-V2X heavily relies on contention-based scheduling due to its autonomous resource selection in Mode 4. This decentralized approach eliminates central arbitration (i.e., flow control) and introduces vulnerability under congested conditions (*6*). In Mode 4, vehicles select communication resources based on local sensing of channel availability, a method known as sensing-based semi-persistent scheduling (SB-SPS) (*6*). While this method allows vehicles to operate independently of cellular infrastructure, it also increases the likelihood of resource collisions, especially when many connected vehicles attempt to transmit BSMs simultaneously in dense traffic scenarios. Without a mechanism to dynamically coordinate or control message rates, excessive legitimate traffic can saturate the channel, leading to increased latency, packet loss, and degraded reliability. This challenges the fundamental performance goals of C-V2X technology, particularly for connected vehicle safety applications that require timely and reliable message delivery (*7*).

To regulate message traffic and mitigate congestion, 3GPP defines Decentralized Congestion Control (DCC) mechanisms based on two channel-level metrics; these metrics are the Channel Busy Ratio (CBR) and the Channel Occupancy Ratio (COR) (*8)*, (*9*). CBR estimates the proportion of busy channels, while COR reflects a transmitter's own usage of a channel. When COR exceeds the predefined threshold relative to the measured CBR, connected vehicles must adjust their message rates and resource usage by dropping or delaying application-layer packets (*7*). However, these congestion controls prioritize radio stability over application responsiveness and may hinder the timely delivery of crucial safety messages (*10*).

Although these mechanisms aim to stabilize Mode 4's decentralized operation, they can be exploited within the bounds of protocol compliance. Rather than relying on spoofing or jamming, protocol-compliant misuse could leverage valid yet abnormal transmissions (e.g., high-rate or oversized payload BSMs) to degrade C-V2X communication performance. These behaviors evade traditional security filters, hence, can overwhelm C-V2X receivers, elevate latency, and suppress safety alerts. In the present study, the dominant degradation is shown to occur at the receiver host/application path, where high-rate but valid traffic increases queue buildup, CPU contention, and FCW processing delay, even while the PC5 safety link remains standards-compliant.

As demonstrated by Trkulja *et al.* (*11*)*,* adversarial selection of resource blocks, while fully compliant with SB-SPS standards, can introduce systematic collisions that significantly reduce packet reception. Moreover, Twardokus and Rahbari (*12*) showed that authentic BSMs transmitted at excessive rates deplete receiver resources, induce delays, and suppress application-layer alerts. Even proposed mitigations like interleaved one-shot SPS in (*13*) still observe meaningful degradation in PDR, inter-packet gap, and age of information. While these studies have shown how standards-compliant message behaviors can cause performance degradation, most have relied on simulation environments or focused on either the transport or the application layer. Moreover, few have evaluated how combined protocol-compliant attacks, involving multiple layers, impact actual C-V2X-based safety applications in real-world conditions.

Connected vehicles' communications are also managed by the IEEE 1609 family, which define the Wireless Access in Vehicular Environments (WAVE) protocol stack (*14*). Positioned above the User Datagram Protocol (UDP) layer, WAVE provides essential services that enable secure, low-latency vehicular communications. A critical component of this family is IEEE 1609.3, which defines networking and transport services for WAVE, including the WAVE Short Message Protocol (WSMP) for low-latency safety messages and full IPv6 support for non-safety applications (*15*). While the IEEE 1609 security framework establishes a robust foundation for trusted communications, it does not fully address the threat of protocol-compliant attacks. Such attacks exploit standard-compliant behaviors, thus evading detection while still degrading the performance of safety-critical applications. These gaps are particularly concerning for C-V2X-based connected vehicle applications.

This paper addresses these gaps by presenting a real-world case study of protocol-compliant Denial-of-Service (DoS) attacks on a C-V2X-based FCW system. We investigate flooding scenarios at both the transport layer (via high-frequency UDP packets) and the application layer (via oversized BSMs), and demonstrate their combined effect on message delivery, latency, and FCW alert suppression. Using a connected vehicle testbed with commercial On-Board Units (OBUs), we conduct empirical evaluations, revealing how standards-compliant, yet excessive traffic can severely degrade safety-critical communications without violating any protocol specifications. This paper makes three key contributions:

- We design and implement a real-world flooding attack that primarily stresses the receiver through high arrival pressure and queue/backlog growth and evaluate its effect on communication reliability and FCW functionality.

- We develop a valid oversized-BSM flooding attack that increases per-message parsing and handling cost at the receiver and analyze how this protocol-compliant behavior affects message processing and safety alert generation.
- We evaluate the combined effect of simultaneous high-arrival-pressure flooding and high-processing-cost flooding, providing an empirical understanding of how these mechanisms jointly degrade C-V2X FCW performance.

The remainder of this paper is organized as follows. First, we provide background on the C-V2X communication stack and related studies on performance degradation in connected vehicle systems. Second, we present the attack model and describe the real-world experimental testbed using commercial OBUs for evaluating protocol-compliant flooding attacks. Third, we detail the attack implementation, including transport-layer UDP flooding, application-layer BSM flooding, and their combined effects. Next, we present empirical results quantifying the impact on message delivery, latency, and FCW alert responsiveness, and we discuss the broader implications for C-V2X safety and resilience. Finally, we present potential mitigation approaches, including the empirical assessment of one mitigation strategy, discuss study limitations, and outline conclusions and future research directions.

## BACKGROUND AND RELATED WORK

This section explains the C-V2X communication stack, IEEE 1609 security framework, and DCC mechanisms relevant to DoS resilience. It also reviews prior research on congestion and protocol-compliant vulnerabilities to present how this study addresses the gap within relevant literature.

### Background

To fully understand how protocol-compliant DoS attacks can impact C-V2X-based safety systems, it is important to first clarify the underlying technical architecture. C-V2X communication technology is organized in a hierarchical stack similar to the Open Systems Interconnection (OSI) model, consisting of four primary layers as illustrated in Figure 2; the physical layer, the Medium Access Control (MAC) layer, the network and transport layers, and the application layer (*16*) , (*17*). Each of these layers performs a specific role, collectively enabling low-latency and reliable communication required by real-time connected vehicle safety applications.

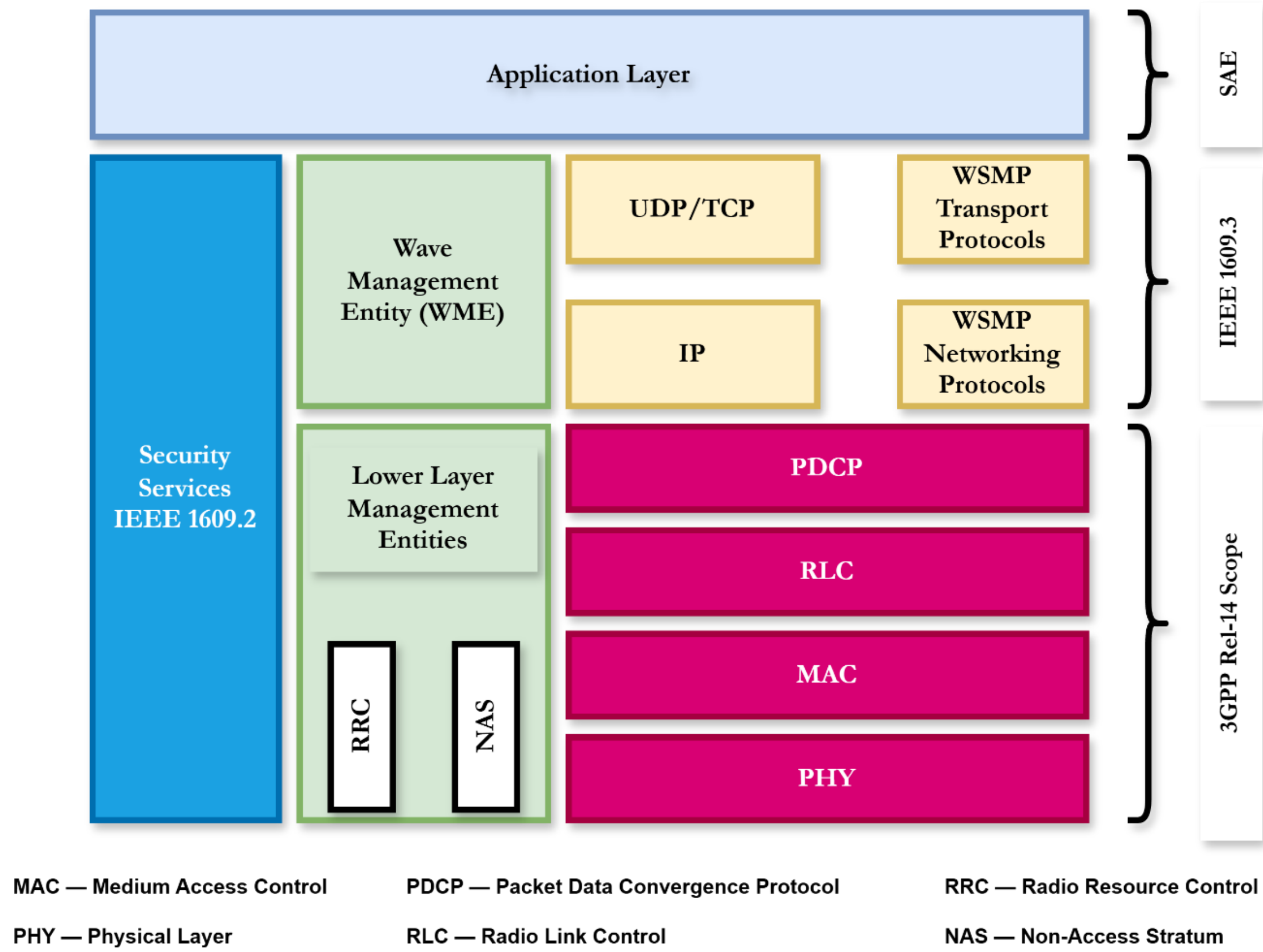

**Figure 2. C-V2X communication stack**

*Physical Layer*

At the physical layer (denoted as PHY in Figure 2), C-V2X functions primarily within the dedicated 5.9 GHz Intelligent Transportation Systems (ITS) band, which has been reserved solely for supporting reliable, short-range vehicular communication for road safety purposes (*18*), (*19*). Communication in this ITS band leverages Orthogonal Frequency-Division Multiplexing (OFDM), a digital modulation technique that divides the available frequency band into multiple orthogonal sub-channels. Across these sub-channels, the OFDM simultaneously transmits signals to effectively mitigate the effect of multipath fading (*20*). Furthermore, OFDM's inherent resilience to interference and its ability to support high data rates and low latency transmission make it particularly suitable for the rapidly changing conditions encountered by vehicles moving at high speeds (*21*).

In the context of our study, the physical layer's design for robust and rapid transmission is critical because it serves as the foundational layer that carries higher-layer messages, such as BSMs, over the air. Although this layer is optimized for reliability and performance, its effectiveness inherently depends on efficient and coordinated use of the wireless channel at the layers above it. Excessive or poorly managed transmissions originating from upper layers (such as the transport and application layers targeted by our DoS scenarios) can severely degrade the physical layer performance. This shows increased interference, reduced signal quality, and diminished reliability of safety-critical communication, ultimately hindering the timely delivery of safety alerts (*7*), (*22*), (*23*).

*MAC Layer*

Directly above the physical layer is the MAC layer, which manages how vehicles access the wireless channel. In C-V2X Mode 4, the MAC uses SB-SPS, allowing vehicles to autonomously select time-frequency slots based on sensed local channel activity. As mentioned in the introduction, this decentralized method enables communication without cellular infrastructure but also increases the risk that multiple vehicles choose the same slot simultaneously, especially in densely populated road environments resulting in packet collisions, interference, and message loss (*1*).

Though SB-SPS typically delivers low latency under normal load, this advantage collapses under congested or adversarial traffic. Multiple vehicles selecting identical resources or receiving excessive, protocol-compliant transmissions can saturate MAC resources, cause scheduling collisions, and impair the timely delivery of BSMs (*24*). In our study, we focus on how upper-layer floods, i.e., deliberate but syntactically valid traffic originating from the application or transport layers, exacerbate vulnerabilities in SB-SPS and compromise real-time safety functions such as FCW.

*Network and Transport Layer*

Above the MAC layer lies the network and transport layers, responsible for ensuring efficient data exchange between vehicles. In C-V2X, the UDP is used on the transport layer due to its minimal overhead and rapid transmission capability. UDP operates on a connectionless basis, meaning it sends packets without requiring acknowledgments or retransmissions, significantly reducing latency (*12*). While ideal for safety-critical messaging, this design inherently lacks congestion control or flow regulation. Consequently, an excessive rate of UDP packets, such as empty packets sent at high frequency, can easily saturate receiver-side buffers, leading to increased packet drop rates, prolonged processing delays, and reduced reliability in message delivery.

This vulnerability directly motivates our research, as we investigate how even protocol-compliant yet high-frequency UDP packet flooding from legitimate nodes can overwhelm receivers and negatively impact the safety applications' function, including collision warning alerts.

*Application Layer*

At the top of the communication stack, the application layer is responsible for formulating and parsing messages that carry critical safety information. In C-V2X deployments, vehicles regularly broadcast BSMs in compliance with the SAE J2735 standard (*25*). These BSMs serve as input for safety applications, such as FCW, EEBL, and intersection movement assist. While BSMs follow standardized formats, optional data fields allow flexibility in message size, potentially creating opportunities for oversized yet valid messages. Transmitting these larger messages at unusually high frequencies can exhaust receiver-side processing resources, delay the parsing and processing of legitimate safety alerts, and ultimately disrupt timely safety responses. This scenario, which we refer to as application-layer flooding, is what our study explores through empirical evaluations, demonstrating how legitimate yet excessive messaging at the application layer can silently degrade critical C-V2X safety operations.

*IEEE 1609 Family and Security Integration*

The foundation of secure communications in connected vehicle networks is based on IEEE 1609 family of standards, which defines the WAVE protocol stack that enables interoperability among On-Board Units (OBUs) and RSUs (*14*). Within this group, IEEE 1609.2 defines the security layer that is primarily responsible for message authentication, digital signatures, and certificate management, while optional encryption is also supported (*15*). These mechanisms help ensure the trustworthiness of safety-critical messages such as BSMs and their protection against spoofing, tampering, or unauthorized access. To Complement these services, IEEE 1609.3 defines networking and transport layer services that operate above UDP/TCP (*14*). It introduces the WAVE Short Message Protocol (WSMP), which enables low-latency dissemination of safety messages by minimizing protocol overhead. In addition, IEEE 1609.3 supports IPv6-based communication for broader IP applications, allowing simultaneous delivery of safety and non-safety messages (*26*). Together, IEEE 1609.2 and 1609.3 provide structured message handling, prioritization, and secure delivery mechanisms critical to vehicular safety systems.

Despite these protections, prior research has identified that the IEEE 1609 family does not inherently address protocol-compliant DoS attacks. For instance, adversaries can exploit legitimate use of UDP, as supported in IEEE 1609.3, to send high volumes of valid messages without violating protocol specifications (*27*). This excessive yet standards-compliant traffic can overwhelm lower-layer resources, disrupt medium access scheduling (*11*), and consequently degrade the performance of C-V2X-based safety applications. Studies have shown that authenticated and encrypted messages, even when fully compliant with IEEE 1609.2 security requirements, can lead to significant congestion and delayed safety alerts if transmitted in high frequency (*28*).

Our work builds upon this standards-based foundation by empirically demonstrating how protocol-compliant message floods can impair safety-critical C-V2X applications. Using commercial OBUs, we evaluate real-world flooding scenarios involving high-frequency UDP packets and oversized BSMs. These attacks leverage mechanisms defined in IEEE 1609.3 and remain fully standards-compliant under IEEE 1609.2 security policies. By analyzing their impact on message delivery, latency, and FCW alert generation, our study highlights an unmitigated vulnerability within the existing WAVE security framework that warrants further attention from the cybersecurity community.

**Related Work**

Previous research on C-V2X communication has primarily focused on the reliability and efficiency of safety-critical messaging under normal and congested network conditions. Studies have extensively explored the resilience of C-V2X against common cyber threats, including spoofing attacks, jamming, and packet manipulation (*21)*, (*29*). However, relatively few studies have investigated how strictly protocol-compliant behavior could become a source of vulnerability.

Early research by Boban *et al.* (2018) and Garcia *et al.* (2021) provided a foundational understanding of the capabilities and limitations of C-V2X, specifically analyzing communication performance metrics, such as latency, PDR, and interference resilience (*30*). These studies revealed the robust nature of C-V2X compared to traditional technologies like DSRC, but primarily addressed ideal or moderately congested scenarios, leaving extreme conditions unexplored. Recent contributions by Twardokus and Rahbari (2022) highlighted potential vulnerabilities coming from legitimate, yet excessive, BSM transmission rates (*12*). Their simulation-based analysis indicated significant receiver-side congestion and processing delays when faced with abnormally high message rates. Although insightful, these simulations lacked validation through real-world deployments and only considered single-layer threats, i.e., either at the transport or the application layer, independently. Fouda *et al.* (2023) expanded on these findings by examining the impact of Hybrid Automatic Repeat Request (HARQ) retransmissions and the resulting delays in safety-critical messages under congested conditions (*31*). Their work suggested that although retransmissions improve reliability, they simultaneously introduced substantial latency, potentially jeopardizing timely safety warnings. Yet, their analysis remained predominantly theoretical and did not examine combined-layer threats. McCarthy *et al.* (2021) developed OpenCV2X, i.e., an open-source simulation environment, specifically modeling the impact of oversized or irregular traffic patterns on sidelink scheduling performance (*32*). Their work demonstrated that irregular application-layer traffic could severely degrade performance metrics. Despite this progress, their simulations did not explore the combined effect of simultaneous transport-layer flooding, nor were the findings validated through field testing.

Prior studies (*11)*, (*25)*, (*32*) has shown through simulation that excessive or irregular traffic can degrade C-V2X communication performance. However, the contribution of the present study is not limited to replacing simulation with field testing. Rather, this work experimentally demonstrates, on commercial Cohda MK6C OBUs running a real FCW application, how different forms of protocol-compliant overload map to distinct receiver-side failure mechanisms and ultimately to application-level FCW outcomes. In particular, the experiments distinguish between lightweight high-rate UDP flooding, which primarily increases arrival pressure and queue/backlog growth, and oversized valid BSM-like flooding, which additionally increases per-packet parsing and handling cost. The study then shows that combining these mechanisms produces the strongest degradation in latency, PDR, and FCW responsiveness.

Consequently, there remains limited empirical evidence regarding how a combined protocol-compliant flooding simultaneously at the transport and the application layers could affect real-world safety-critical applications such as an FCW system. Our study addresses this gap by conducting a real-world testbed-based experimental evaluation of such multi-layer DoS scenarios. We measure real-time impacts on latency, message delivery, and FCW application performance, offering practical insights and present corresponding mitigation recommendations that do not require changes to existing C-V2X standards or protocols.

## METHODS

The methods section describes the experimental approach used to evaluate DoS threats. It explains the attack model, experimental testbed, and metrics used to assess performance degradation under flooding attacks.

### Experimental Method

To assess the resilience of C-V2X FCW systems under realistic adversarial conditions, we devised an experimental method, as shown in Figure 3. For these experiments, we first define a realistic attack model in which an adversary uses a fully compliant C-V2X device to generate excess traffic at both the transport and the application layers. Next, we establish baseline performance by driving two vehicles equipped with commercial Cohda MK6C OBUs under normal conditions, measuring PDR, end-to-end latency, and FCW trigger timing. Finally, we implement and execute protocol-compliant flooding attacks, transport-layer UDP flooding, and application-layer oversized BSM floods, both independently and in combination while continuously monitoring the same performance metrics. By comparing attack scenarios against our baseline, we quantify how each flooding strategy degrades communication reliability and delays or suppresses collision warnings, thus revealing the vulnerabilities of C-V2X FCW systems under stealthy, standards-compliant DoS conditions.

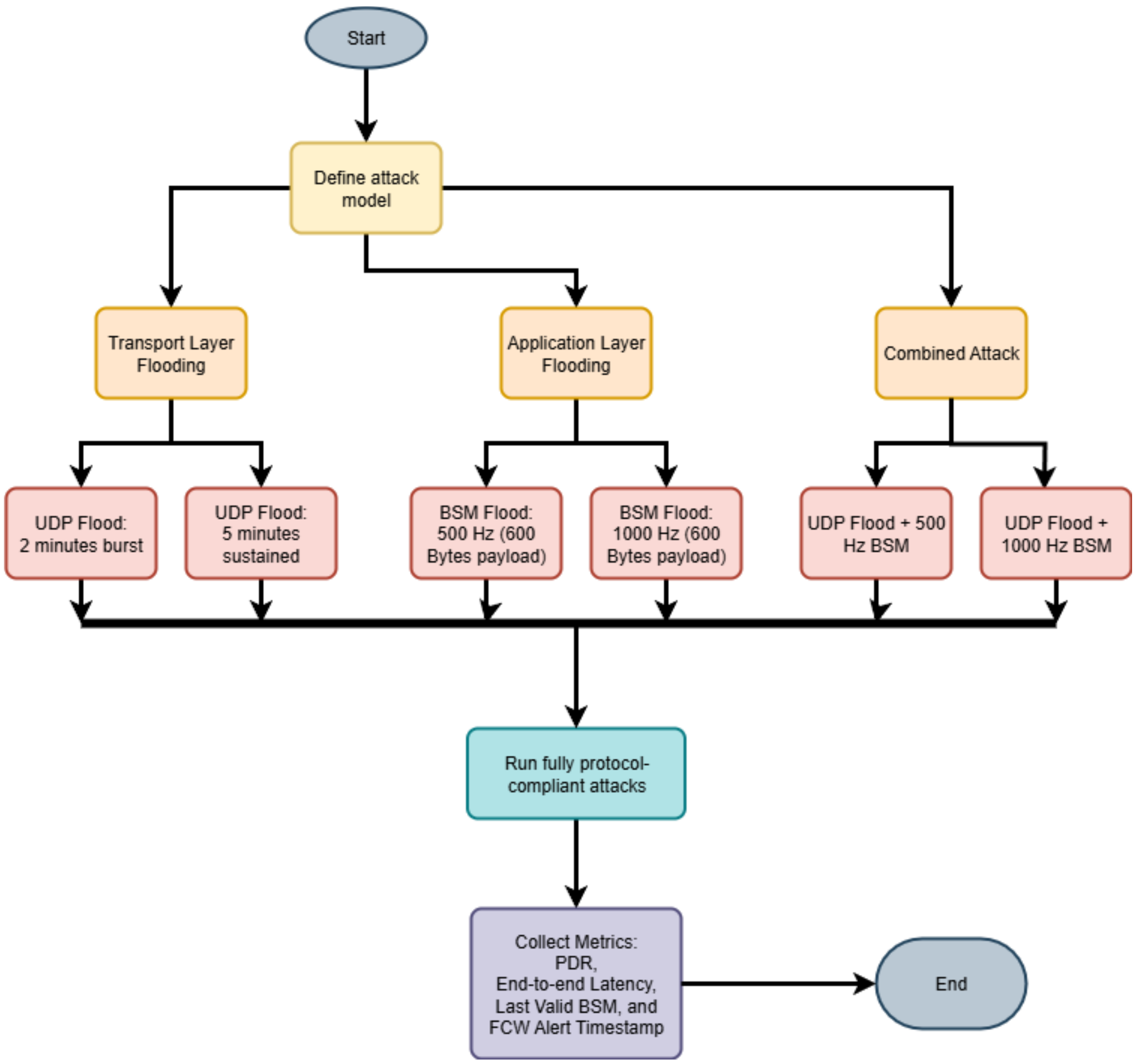

**Figure 3. Experimental plan for three DoS attack scenarios**

**Attack Model**

In our attack model, we consider an adversary with full access to a standards-compliant C-V2X device, including the capability of transmit BSMs and UDP packets over the PC5 sidelink. Importantly, the attacker does not spoof message content, manipulate identifiers, or violate protocol specifications. Instead, the adversary exploits a design-level weakness in C-V2X communications: the lack of built-in rate control and congestion management mechanisms.

By deliberately increasing the volume and frequency of otherwise valid transmissions, the attacker can saturate wireless channel resources, overload receiver-side message queues, and delay the timely processing of genuine safety-critical data. Because these transmissions fully comply with the 3GPP and SAE J2735 standards, they are difficult to distinguish from legitimate traffic. This makes the attack stealthy and challenging to detect using conventional intrusion prevention or anomaly-based filtering systems.

This model reflects a realistic and practical attack scenario, in which a malicious, compromised, yet fully authorized vehicle or RSU generates excessive, protocol-compliant traffic. Such behavior could occur unintentionally due to faulty firmware or can be exercised maliciously due to adversarial reprogramming. In either case, the result is a DoS condition that silently degrades communication reliability and delays safety application responses without breaching security policies. As mentioned before, we target two layers in this attack model as follows:

- **Transport Layer**: Exploiting UDP's connectionless and unregulated transmission behavior to create floods that saturate receiver buffers.
- **Application Layer**: Transmitting oversized or abnormally frequent BSMs that adhere to protocol rules but induce excessive processing delays and resource contention.

The objective of the adversary is to degrade the performance of a C-V2X-based FCW system while remaining fully compliant with protocol specifications. Specifically, the attacker aims to reduce packet delivery and increase end-to-end latency to levels that compromise safety alert generation. We evaluate this degradation using the following performance metrics:

$$PDR = \frac{N_{recv}}{N_{sent}} \times 100\% \tag{1}$$

$$L_{avg} = \frac{1}{N_{recv}} \sum_{i=1}^{N_{recv}} (t_i^{rx} - t_i^{tx}) \tag{2}$$

where, $N_{sent}$ and $N_{recv}$ are the number of BSMs transmitted and successfully processed by the FCW logic, respectively, and $t_i^{rx}$, $t_i^{tx}$ are the transmission and reception timestamps for message *i*.

Under normal conditions, these metrics satisfy application-layer requirements (i.e., PDR ≥ 90% and $L_{avg} \leq$ 50ms). However, when the aggregate arrival rate at the receiver exceeds the host/application dispatch capacity, the receiver message queue becomes overloaded. This leads to increased delay and packet loss, which can be modeled as:

$$Q(t + \Delta t) = Q(t) + A(t) - D(t) \tag{3}$$

where $Q(t)$represents the instantaneous queue length (i.e., the number of unprocessed messages) at time $t$, $A(t)$is the aggregate packet arrival rate including both benign and adversarial traffic, and $D_{eff}(t)$denotes the effective receiver-side dispatch rate supported by the available host/application processing capacity. Under normal conditions, $A(t) = D_{eff}(t)$maintains a stable queue state, while $A(t) < D_{eff}(t)$causes the queue to shrink because messages are processed faster than they arrive. In contrast, when $A(t) > D_{eff}(t)$, the queue grows over time, leading to backlog accumulation, buffer overflow, increased latency, and possible suppression of safety-critical alerts such as FCW.

In the application-layer flooding scenario, the attacker transmits oversized yet syntactically valid BSMs. Although these messages do not necessarily exceed the physical channel capacity, they impose higher computational cost on the receiver. Larger payloads require additional parsing, validation, and memory operations for each message, which reduces the number of messages that can be processed within a given time interval. This effect can be expressed as

$$D_{eff}(t) = \frac{1}{T_{processing}(S)} \tag{4}$$

where $T_{\mathrm{processing}}(S)$is the average per-message processing time and increases with payload size $S$. Therefore, even when the offered traffic remains within normal physical channel limits, the effective dispatch rate can still decrease because larger packets consume more host-side processing resources. As a result, queue buildup and latency inflation may still occur due to overload at the receiver host/application path rather than intrinsic saturation of the PC5 radio interface alone.

## CASE STUDY: PROTOCOL-COMPLIANT DOS ATTACKS IN CONNECTED VEHICLES

This section details the real-world implementation of the attack model using two connected vehicles and commercial OBUs. It establishes baseline FCW performance, describes the staged DoS attack executions, and explains how each scenario stresses system resources.

### Experimental Setup and Baseline Operating Condition

We evaluate the safety application, FCW, under our attacks in a real-world testbed, as shown in Figure 4. Using two vehicles, each equipped with a Cohda MK6C Evaluation Kit (EVK) OBU running 3GPP Release 14 firmware and a rooftop omnidirectional antenna for robust sidelink communication. Figure 4 illustrates the experimental setup where vehicle A moves towards vehicle B under normal driving conditions.

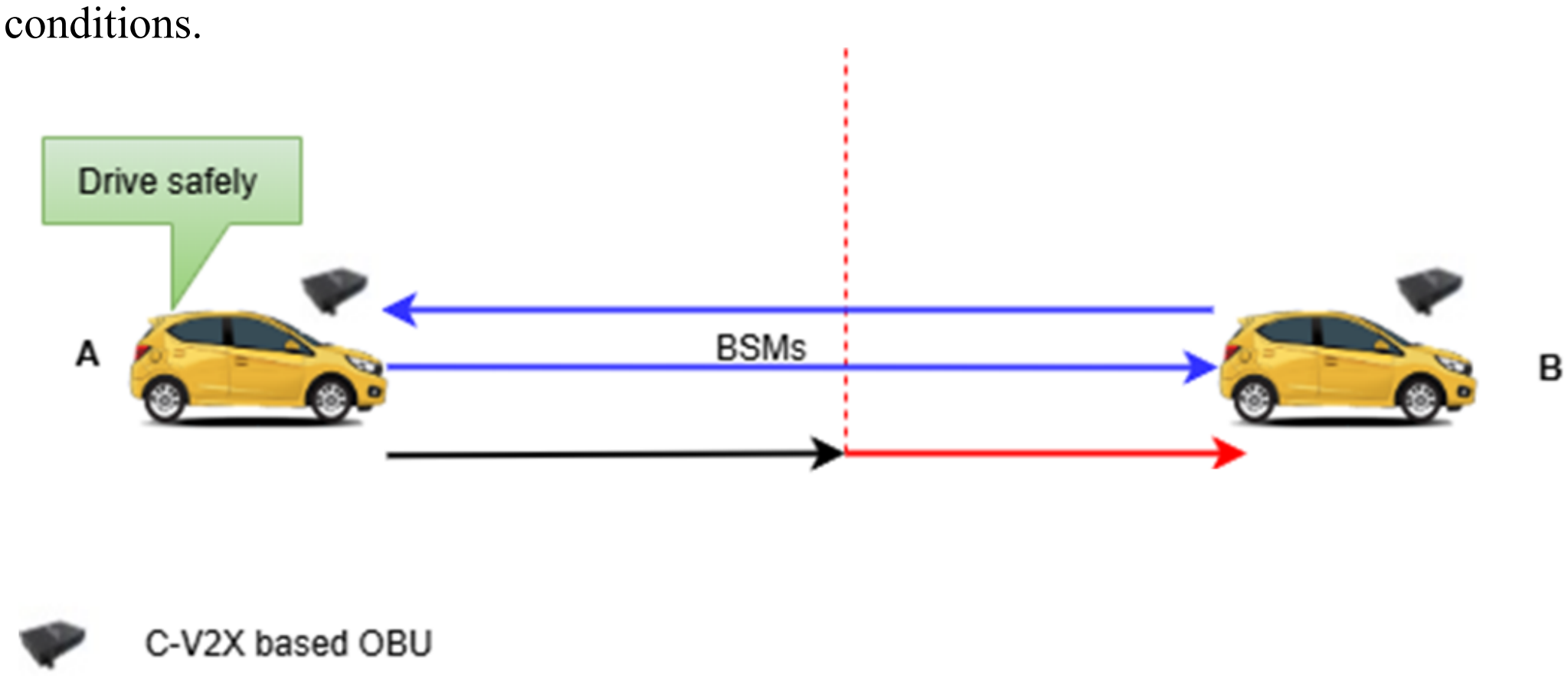

**Figure 4. Experimental setup**

Each OBU broadcasts and receives BSMs at a frequency of 10 Hz. In this study, we used the BSM fields speed, longitude, latitude, and braking status, as these are essential inputs for the FCW algorithm to evaluate collision risk and trigger timely alerts. The FCW application continuously computes the Time-to-Collision (TTC) between two vehicles using their relative positions and speeds:

$$TTC\ (t) = \frac{d(t)}{v_A(t) - v_B(t)} \tag{6}$$

where, $d(t)$ is the gap between vehicles A and B, and $v_B(t)$, $v_A(t)$ are the speeds of vehicles A and B, respectively.

In our, as illustrated in figure 5, vehicle B is stationary (i.e., $v_B(t) = 0$) and the follower vehicle A is moving, therefore the *TTC* simplifies to $TTC\ (t) = \frac{d(t)}{v_A(t)}$ and a safety alert is issued when the *TTC* falls below a threshold of 3 seconds, minimum to balance timely warning and false-alarm rate in FCW systems (*33*).

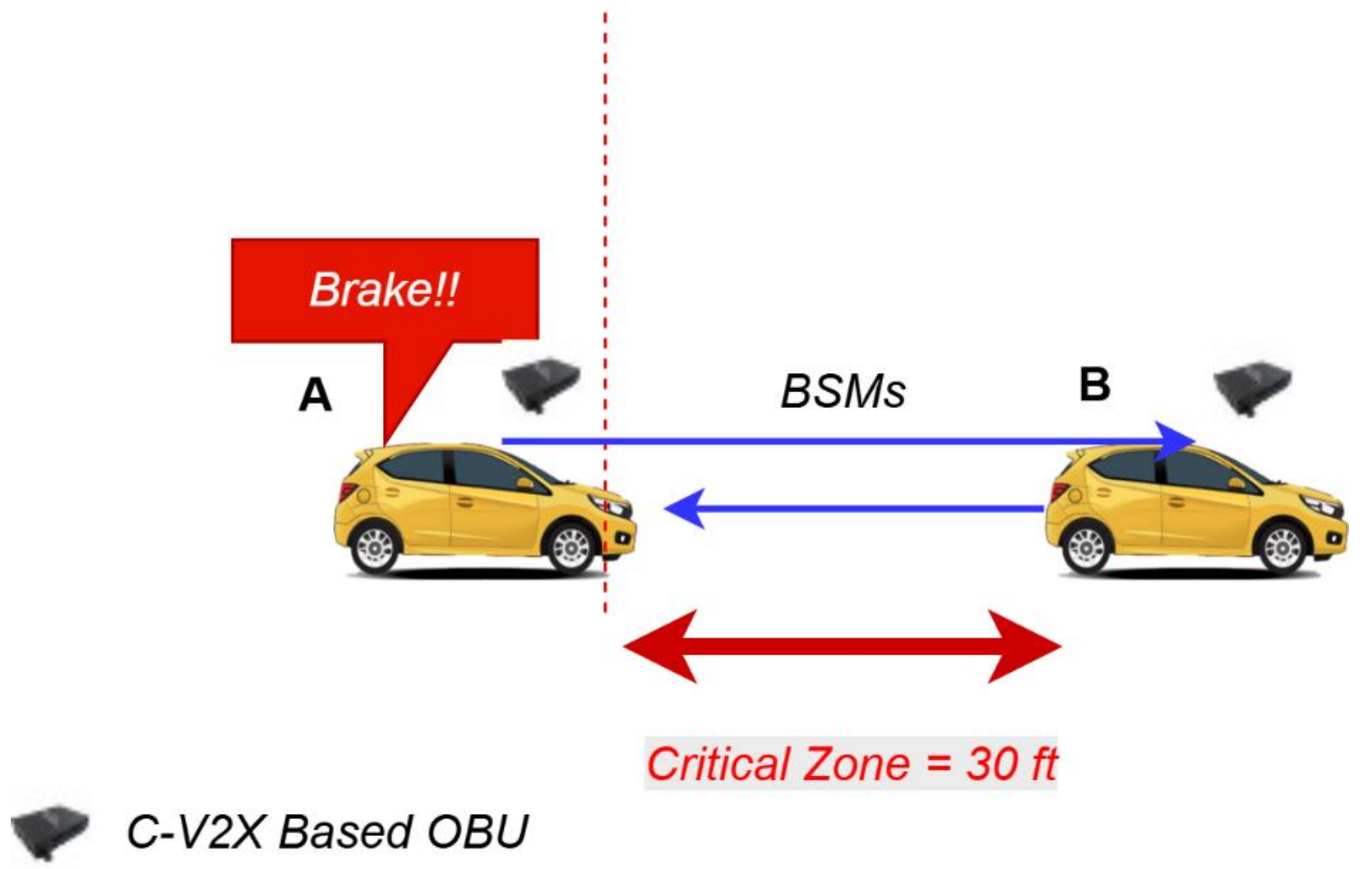

**Figure 5. FCW alert when TTC falls below the threshold chosen for this study**

Before introducing the DoS attacks, we established a baseline to define the normal performance of our connected vehicle FCW application. At normal conditions, the system operates at a 10Hz BSM transmission rate, and the C-V2X PC5 sidelink maintains high reliability and low latency, ensuring the timely delivery of BSMs for the FCW application.

Figure 6(a) shows the BSM latency distribution over time during our baseline operation. The latency value remained consistently below the threshold of 50 milliseconds, averaging 35 milliseconds. These characteristics align with the C-V2X design target for safety applications and enable the FCW algorithm to continuously compute TTC without delay, triggering alerts promptly when the oncoming vehicle A is within 30 feet of vehicle B, the critical zone. Meanwhile, Figure 6(b) shows the BSM packet rate measured during our baseline operation. The transmission frequency remained stable at 10 packets per second, with minimal fluctuation. This stability contributed to a PDR exceeding 99%, confirming that the communication stack functioned optimally under normal conditions. These baseline performances from our testbed serve as a reference point for the following experiments involving transport and application-layer flooding attacks.

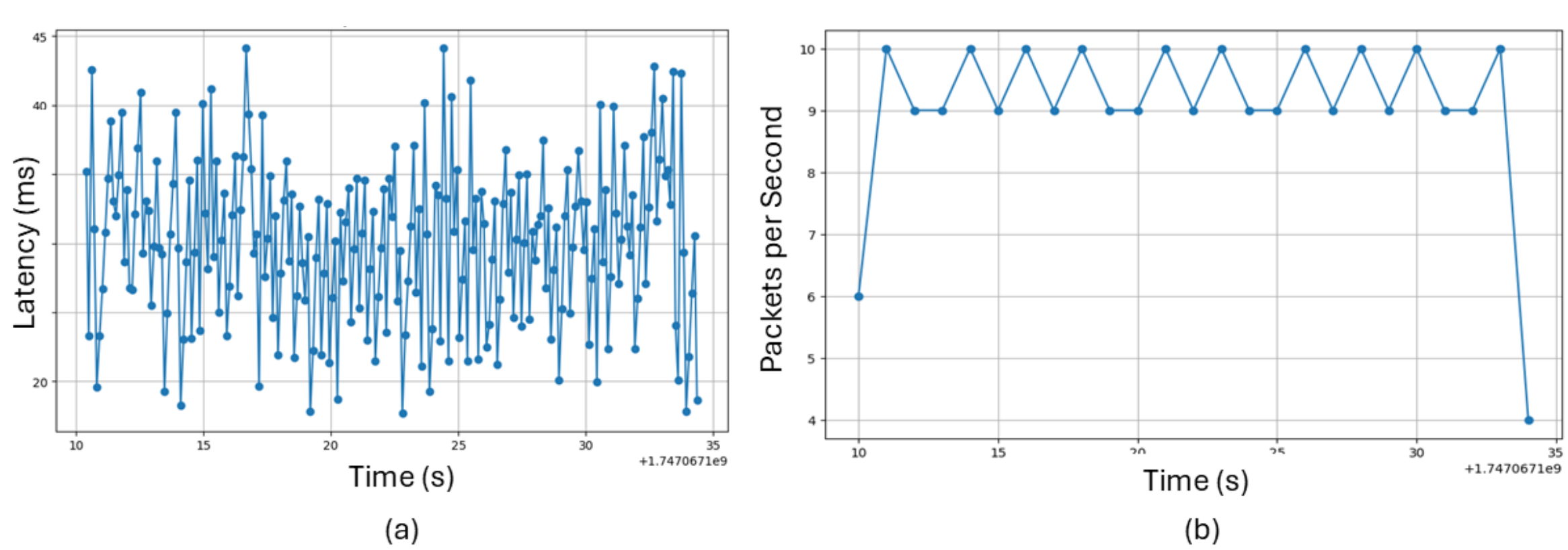

**Figure 6. C-V2X communication performance for baseline (i.e., non-attack) operating condition**

**Attack Execution**

Building on the experimental setup described above, we designed and executed a series of DoS attacks to evaluate the impact of protocol-compliant flooding on the FCW application. The same two-vehicle testbed equipped with Cohda MK6C OBUs was used and we did not use the Application Program Interface (API) from the Software Development Kit; however, as shown in Figure 7, we introduced an additional attack control system acting as an adversarial node. This system consists of a high-performance laptop directly interfaced with one of the OBUs, allowing us to generate synthetic yet standards-compliant traffic. This approach effectively transforms the OBU into a malicious connected vehicle node without requiring hardware modifications or protocol violations. The node was controlled via Linux Secure Shell (SSH) and executed Python 3 scripts to produce a minimal-payload UDP flood, and a high-rate, oversized BSM-like JSON stream over UDP; see Appendix A, Algorithms A-1 and A-2 for implementation details and parameters. The attacker host/OBU remained stationary, positioned at 80 ft from Vehicle B (receiver) with direct line-of-sight. Adversarial traffic was delivered to the receiver's non-safety UDP/IP ingress; the PC5 safety path was not used for attack traffic and remained standards-compliant throughout.

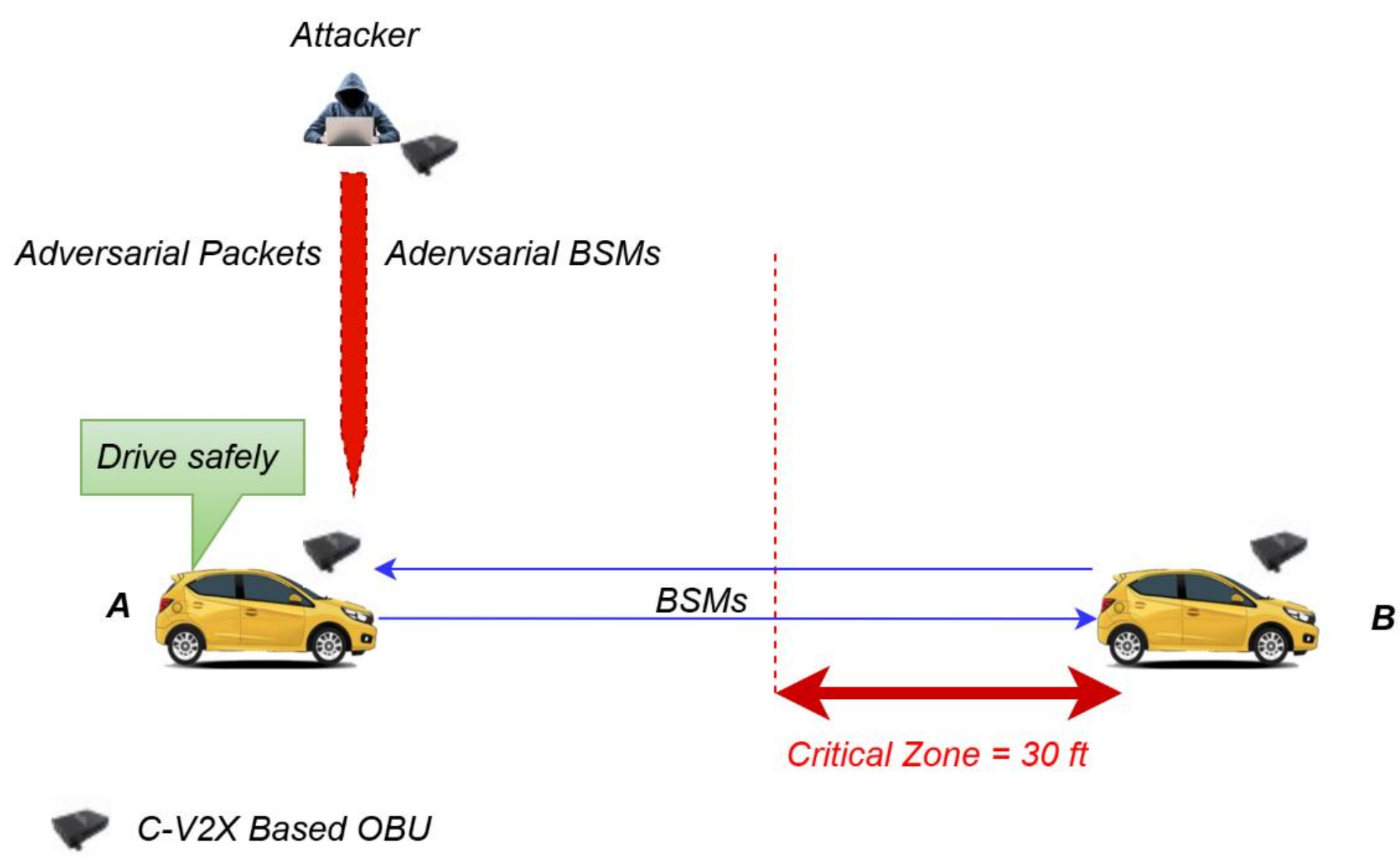

**Figure 7. Attack setup**

The first wave of attacks targeted the transport layer. Consider a scenario where an otherwise normal device suddenly starts sending an overwhelming number of empty UDP packets. The messages themselves are valid and comply with communication protocols, but the sheer volume quickly overwhelms the receiver's buffer. We tested this behavior in two ways. In the first case, the attacker unleashed a short yet intense burst lasting two minutes, similar to a quick, disruptive wave. In the second case, the flood was kept for five minutes, persistently filling message queues and forcing the system to struggle with processing delays. Both attacks were fully standards-compliant, which made them appear legitimate, yet they were specifically designed to exhaust resources and slow down the flow of safety-critical information to the FCW system.

After observing how the transport layer responded, we crafted attacks aimed at the application layer. Normally, vehicles exchange BSMs at 10Hz, which is a balanced rate designed for reliable and timely communication. For this application-layer attack, the attacker drastically increased this rate to 500 BSMs per second, overwhelming both the channel and the receiver's ability with a flood of messages. While every message still followed protocol rules and appeared legitimate, the excessive volume congested the communication pathway. Processing legitimate safety messages became a challenge. To push the system further, we doubled the rate to 1,000 messages per second, overwhelming the receiver's queueing and processing capacity and extending processing resources to their limits.

Finally, to replicate more real-world attacks, we combined these flooding techniques. In one scenario, the system faced a UDP flood and a simultaneous 500 Hz BSM with a payload of 600 bytes, creating two simultaneous stressors: high arrival pressure from the UDP flood and increased per-message processing burden from the oversized BSM-like flood. The ultimate test came when we paired a sustained UDP flood with a 1,000 Hz BSM, a worst-case attack that saturated communication buffers and message handling capabilities, leaving the system on the edge of a complete communication breakdown.

These attacks were carefully crafted to look normal while behaving stealthily enough to bypass conventional detection. By gradually escalating the attack duration, message frequency, and layering techniques, we crafted realistic adversarial conditions that exposed how protocol-compliant message floods could silently disrupt communication and delay safety-critical alerts of an FCW application.

Table 1 illustrates the experimental environment where all trials were conducted on a controlled closed track with no public traffic to minimize external variables. Only the two instrumented vehicles were present in the radio neighborhood; Vehicle B (benign transmitter) was stationary, and the attacker node was also stationary at a fixed track location. The on-air PC5 sidelink and standards-conformant safety messaging were preserved; we stressed only the receiver's non-safety UDP/IP ingress using (A) a minimal-payload UDP flood and (B) 600-byte "BSM-like" JSON stream at 500/1000 Hz for 120s/300s, while benign FCW messages continued at 10 Hz. Runs were scheduled in off-peak windows, under stable weather, and we qualitatively observed a quiet RF environment along the track.

**Table 1. Environment Controls**

| Factor | Setting |
|---|---|
| Roadway context | Controlled closed track; no public traffic. |
| Vehicle and attacker | attacker node stationary (~80 ft from Car 2(receiver); line-of-sight; traffic injected via non-safety UDP/IP (not PC5).<br>Role of vehicle B: vehicle B (benign) is stationary, sending BSMs to vehicle A.<br>Role of vehicle A: vehicle A (victim) drives and runs the FCW from receiving BSMs from vehicle B. |
| Non-safety ingress | UDP/IP to receiver application (attacks applied only). |
| Traffic density | Only two connected vehicles in a controlled environment. |
| Weather | Not windy, sunny and clear visibility. |
| Operating vehicle speed | Operated within posted track limits (i.e. 25 mph). |
| RF notes | No intentional interferers; qualitatively low background activity. |

## RESULTS

In this section, we present the impact of protocol-compliant flooding attacks on C-V2X-based OBU communications and their direct effect on FCW. Figure 8 and Table 2 present detailed performance metrics for each tested scenario, including PDR, average end-to-end latency, and the resulting impact on FCW alert generation.

Under baseline conditions (10 Hz BSM), the system maintained a high PDR of 99.2% with an average latency of only 35 ms. Consequently, the FCW system reliably triggered timely safety alerts, with the last valid BSM received at 17.00 s and the alert triggered shortly after at 17.42 s. These metrics, i.e., last valid BSM's timestamp and FCW trigger timestamp, are critical for evaluating system responsiveness. The last valid BSM's timestamp represents the final successfully processed safety message before a potential collision, whereas the FCW trigger timestamp indicates when the collision warning is activated based on these messages. These metrics directly reflect the system's ability to issue alerts within the safety-critical TTC threshold of 3 seconds.

These measured trends are also consistent with the analytical latency and packet-loss model developed earlier in the paper. In particular, the model predicts that once the aggregate arrival rate at the receiver exceeds the host/application processing capacity, queue buildup begins to occur, which increases delay and reduces the probability that legitimate packets are processed in time. The model also predicts that this effect becomes more severe when packets require more processing effort, such as larger or oversized payloads. This is exactly the pattern observed in our experiments. The UDP flooding cases primarily increase arrival pressure and drive backlog growth, whereas the oversized but valid BSM flooding cases further reduce the effective processing capacity of the receiver by increasing parsing and handling cost per packet. When both effects are combined, the degradation becomes more severe because the receiver experiences both higher arrival pressure and higher per-packet processing burden at the same time. Therefore, the measured FCW failures are consistent with the analytical prediction that overload in this setting is governed mainly by the receiver host/application path, rather than by saturation of the PC5 radio interface alone.

our experiments show that protocol-compliant flooding gradually exceeds the receiver's host/application processing capacity, which in turn prevents the system from handling new legitimate safety messages on time. Under moderate attack pressure, the receiver still processes some valid BSMs, but the added delay is large enough to cause late FCW warnings. Under more sustained or combined flooding, the processing backlog appears much earlier, legitimate BSMs are dropped sooner, and the FCW system begins to miss alerts entirely. This progression is directly reflected in the last-valid-BSM times and FCW-trigger timestamps reported in Table 2.

When we executed a transport-layer UDP flood attack for 2 minutes, the latency increased to 156 ms and PDR dropped to 89% as shown in Figure 8, causing a delayed FCW alert. The last valid BSM in this scenario was received at 17.21 s, with the alert triggered at 18.30 s. Extending the flood duration to 5 minutes severely impacted system performance, reducing PDR to 12.4% and elevating latency to 400 ms. As a result, the FCW system failed to issue an alert, as the last valid message was received at 5.37 s.

In application-layer flooding scenarios, transmitting valid but oversized BSMs at 500 Hz moderately degraded performance, lowering PDR to 67.3% and raising latency to 105 ms. The FCW alert was triggered at 19.50 s, nearly 2 seconds after the last valid BSM was received at 17.50 s. Increasing the frequency to 1000 Hz intensified these effects, further reducing PDR to 41.5% and increasing latency to 180 ms, resulting in a missed FCW alert.

The most severe impacts occurred under combined flooding attacks. With simultaneous UDP flood and a 500 Hz BSM flood, PDR dropped sharply to 28.7% with latency reaching 250 ms. The last valid BSM was received at 5.12s, meaning the system lost communication significantly earlier than in other "missed alert" cases, leading to a severe impact despite similar alert outcomes. Combining a UDP flood with a 1000 Hz BSM flood caused the most extreme degradation, with PDR falling to 10.2% and latency exceeding 420 ms. The last valid message arrived at 4.83 s, completely disabling the FCW system well before reaching the 3-second TTC threshold, leaving no opportunity for late FCW alert generation.

Collectively, these empirical results show that layered, protocol-compliant flooding attacks substantially compromise both communication reliability and application-layer responsiveness. Thus, the principal effect of protocol-compliant flooding in this testbed is not merely increased traffic volume, but the suppression of timely processing of new legitimate safety messages at the receiver. Earlier breakdowns indicate heavier attack impact, as they eliminate even the possibility of a delayed FCW response.

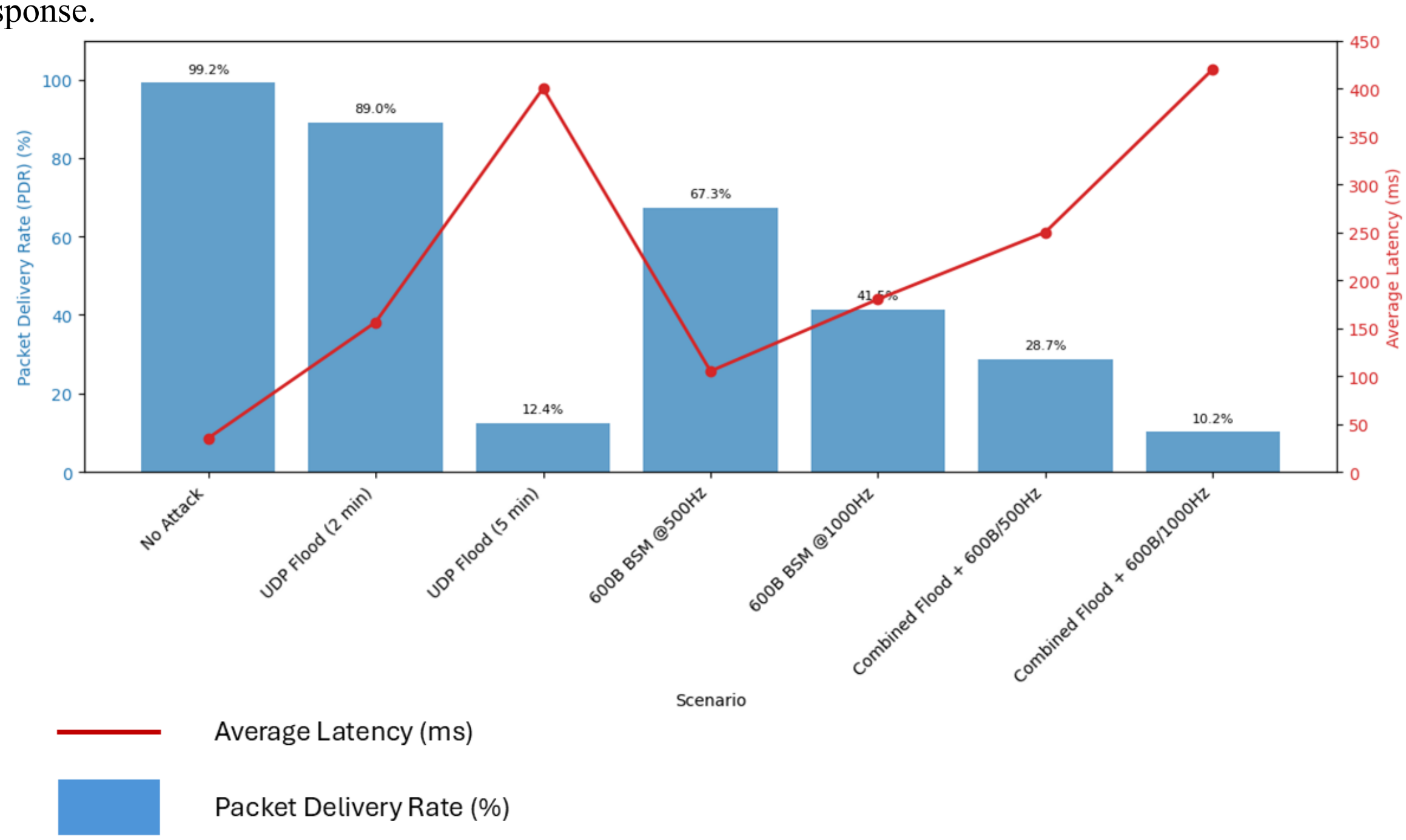

**Figure 8. Impact of transport and application-layer attacks on C-V2X communication performance**

**Table 2. Impact of Protocol-Compliant Attacks on Valid BSM Reception and FCW Functionality**

| Scenario | Average Latency (Maximum Threshold: 50 ms) | PDR (Minimum Threshold: 90%) | Last Valid BSM's Timestamp | FCW Trigger Timestamp | FCW Alert | Attack Success |
|---|---|---|---|---|---|---|
| Baseline | 35 ms | 99.2 % | 17.00 s | 17.42 s | Timely alert | No attack |
| UDP Flood (2 minutes) | 156 ms | 89.0 % | 17.21 s | 18.30 s | Delayed alert | Successful |
| UDP Flood (5 minutes) | 400 ms | 12.4 % | 5.37 s | No trigger | Missed alert | Successful |
| 500Hz Valid BSM Flood | 105 ms | 67.3 % | 17.50 s | 19.50 s | Delayed alert | Successful |
| 1000Hz Valid BSM Flood | 180 ms | 41.5 % | 4.83s | No trigger | Missed alert | Successful |
| Flood + 500Hz BSM (600 Bytes) | 250 ms | 28.7 % | 5.12 s | No trigger | Missed alert | Successful |
| Flood + 1000Hz BSM (600 Bytes) | 420 ms | 10.2 % | 4.83 s | No trigger | Missed alert | Successful |

## DISCUSSION

Our experimental findings reveal a critical vulnerability in current C-V2X safety systems. The ability to severely degrade FCW performance through protocol-compliant attacks demonstrates that adherence to communication standards alone is insufficient to guarantee safe operation of connected vehicle systems. Also, the current design of DCC, while effective at protecting radio interface stability, creates vulnerabilities at higher protocol layers, leaving FCW and similar safety systems exposed to stealthy, standards-compliant DoS conditions.

While this study focused on FCW, the protocol-compliant flooding techniques demonstrated here could be extended to evaluate other critical C-V2X safety applications. EEBL systems, which rely on rapid BSM processing to detect emergency braking events from surrounding vehicles, would likely exhibit similar vulnerabilities to oversized BSM floods that delay message parsing. Also, multi-vehicle attack scenarios could represent a particularly concerning extension of this work. If multiple compromised vehicles in a traffic area simultaneously execute coordinated UDP and BSM floods, the resulting network congestion could effectively disable C-V2X safety communications across an entire road segment.

Although our closed-track study used two vehicles, the protocol-compliant non-safety UDP/IP workload, which contended with the FCW decision loop, can be extended to denser connected vehicle environments. Note that the primary bottleneck we target in our attack experiments is not the PC5 radio link. The bottleneck is the host application path, where the receiver handles non-safety UDP/IP traffic and where valid (i.e., protocol-compliant) traffic consumes CPU and fills queues, delaying the FCW decision loop. Independent resources show that, at the PC5 sidelink, C-V2X is engineered to operate under dense conditions with congestion control. For instance, the 5G Automotive Association (5GAA) Functional & Performance Test Procedures define congestion test methods and demonstrate lab and field behavior; these assess sidelink reliability under standardized congestion control (e.g., SAE J2945/1 and J3161/1) across dozens to hundreds of devices or emulated stations, providing a baseline of radio robustness under load (*34*). Moreover, Qualcomm's C-V2X Congestion Control Study, which tested with a large number of emulated connected vehicles and validated through field trials, showed that the sidelink maintains reliability when rate and admission control are active (*35*). Both sets of experiments include scenarios ranging from 0 to 250 connected vehicles in one-lane and 0 to 500 vehicles in two opposing lanes within a 500-meter PC5 sidelink communication range. Thus, public PC5 performance-related studies indicate the sidelink can sustain high-density loads; therefore, our two-vehicle measurements should be read as a lower-end estimate of host-application delay. As the number of legitimate connected vehicles increases, the benign processing load grows due to the need to parse more BSMs, handle more socket interruptions, and maintain longer receive queues. If the host does not prioritize resources for safety, the extra legitimate work leaves less CPU capacity and buffer headroom. Consequently, protocol-compliant non-safety traffic (including our attacker stream) intensifies contention for CPU cycles and queue space, which pushes the FCW loop later. This scaling is mechanism-based (e.g., congestion control), not radio-limited; the PC5 sidelink can remain standards-compliant under congestion control, yet the application path still slows as aggregate arrival rate increases. Therefore, the attack we report in our experiment should be generalizable to higher-density connected vehicle settings and, in fact, more connected vehicles can amplify the attacker's impact by adding background load that compounds the attack messages' rate-controlled arrivals, unless deployments enforce mitigations, such as ingress monitoring, safety-first scheduling, and payload parsing.

While the experiments in this study were conducted in a controlled two-vehicles environment to isolate the effect of protocol-compliant flooding on the receiver, the identified degradation mechanism is expected to remain relevant in denser traffic or at higher CV penetration if the attack traffic reaches the host/application path. In higher-density traffic, the receiver must process more legitimate neighboring messages in addition to any malicious traffic, which reduces available CPU processing capacity and can increase queue occupancy, backlog, and FCW processing delay. Similarly, if multiple compromised vehicles transmit simultaneously, the aggregate arrival pressure at the victim receiver would increase further, which would likely intensify the overload observed here. Consequently, these results should not be interpreted as a direct quantitative prediction for all deployment conditions. Urban RF environments introduce additional factors such as non-line-of-sight propagation, fading, shadowing, and stronger channel interference, all of which can further degrade the packet arrival process at the host.

The attack behaviors studied here also suggest several practical detection signatures for protocol-compliant flooding. Because the malicious packets remain syntactically valid, detection cannot simply rely only on format checking or standard-compliance tests. Instead, monitoring should also focus on behavioral anomalies such as abnormal per-source ingress rate, sustained BSM transmission frequency above expected norms, repeated oversized yet valid payloads, rising socket backlog, elevated CPU utilization, increased parsing cost, and FCW loop latency excursions. Another useful indicator in this setting is divergence between radio-side and host-side behavior: the radio interface may continue to appear standards-compliant while the receiver's non-safety ingress and application path show increasing queue growth and processing delay. Based on these observations, such signatures could support lightweight threshold-based or anomaly-based monitoring on the OBU and could be used to trigger defensive actions such as source throttling, queue isolation, watchdog intervention, or host-side such as token-bucket policing and FQ-CoDel.

Furthermore, this work intentionally focuses on protocol-compliant flooding attacks, as opposed to other forms of cyberattacks, such as spoofing and jamming, which have already been extensively investigated in prior research. By narrowing the scope, this study isolates and highlights the often-overlooked threat of compliant flooding attacks. However, future research could combine these well-studied attacks with protocol-compliant threats to provide a more holistic understanding of security challenges in C-V2X networks. Additionally, the hardware and software platform used in these experiments may not fully represent the behavior of other commercial OBUs or safety application architectures, potentially influencing latency and packet delivery outcomes. Finally, the analysis centered on FCW alone, while other safety-critical applications that rely on timely message delivery, such as cooperative braking or intersection collision avoidance, were not inspected.

## POTENTIAL MITIGATIONS

Radio-layer congestion controls such as SAE J2945/1 for vehicle-to-vehicle safety, SAE J3161/1 (*34*) for LTE-V2X, and the European Telecommunications Standards Institute (ETSI) Decentralized Congestion Control (DCC) help keep the PC5 sidelink timely at high neighbor density; however, they do not regulate admission or prioritization once packets arrive off radio at the host. Our results show that valid UDP/IP traffic can still inflate FCW processing time by competing for CPU and socket queues, even when the PC5 interface is operating within specification. Therefore, C-V2X deployments require both layers: Radio Frequency (RF)-side congestion control and host/application protections for which we recommend the following measures:

**Ingress monitoring and low-latency queuing:**

In practice, ingress monitoring should be enforced at the OBUs via per-source token-bucket limits (Linux traffic control) and paired with a low-latency queuing discipline such as Fair Queuing Controlled Delay (FQ-CoDel) to cap abusive arrivals and bound queue job (*36*). In our setting, these controls reduced tail delay and improved resilience under both the UDP-flood and 600-byte BSM-shaped JSON workloads, as quantified in Figures 9–11**.**

**Strict priority for safety processing**:

In parallel, safety processing should receive strict priority. Specifically, pin FCW threads to dedicated CPU cores and elevate their scheduling class to real-time and separate receive buffers/queues so safety packets bypass non-safety backlog. Together, this host-side isolation complements J2945/1 and DCC by protecting the compute path after packet reception (*37*).

**Admission control and source isolation:**

Admission control and source isolation should further whitelist expected non-safety peers and sandbox all others in containers or namespaces with Linux control groups (cgroups) enforcing CPU and Input/Output allocations, preventing standards-compliant but abusive senders from monopolizing resources (*38*), (*39*).

**Telemetry with watchdog actions***:*

Finally, telemetry with watchdog actions should continuously track ingress rate, socket backlog, CPU utilization, and FCW loop latency, triggering controlled degradation (e.g., shedding non-critical analytics or tightening policing) when thresholds are exceeded (*34)*, (*40*).

Together, these measures could address the application-path resource contention evidenced in our experiments while remaining fully compatible with PC5-sidelink congestion control.

**Evaluation of the Token-Bucket and FQ-Codel**

Because our analysis identified the dominant bottleneck at the receiver host/application path, rather than at the PC5 radio interface, we evaluated a practical host-side mitigation using per-source token-bucket policing together with FQ-CoDel on the receiver's non-safety UDP/IP ingress. The token bucket limits abusive per-source arrival bursts before they consume excessive CPU time and socket-buffer resources, while FQ-CoDel helps control standing queue growth and reduce tail latency.

Figure 9 compares measured communication performance with this mitigation disabled and enabled under the same attack scenarios. Under benign conditions, mitigation preserves performance close to the no-attack baseline (i.e., mitigation OFF under no-attack case), indicating that it introduces only limited disruption during normal operation. Under attack, the mitigation improves resilience across all evaluated cases, but the pattern of improvement also helps explain the underlying overload mechanism. In the shorter UDP flooding case (i.e., the 2-minute UDP flood condition), the mitigation is comparatively effective because early rate limiting and queue control are sufficient to keep the receiver from entering deep backlog, allowing legitimate packets to continue being processed with only moderate disruption. However, as the overload becomes more sustained (i.e., as attack exposure duration or combined attack pressure increases), the benefit decreases. In the longer-duration UDP flood (i.e., the 5-minute UDP flood condition), and especially in the combined flooding cases (i.e., simultaneous UDP flooding with 600-byte BSM flooding at 500 Hz or 1000 Hz), the receiver remains under substantial stress even after high-rate packet arrivals are partially constrained, so queueing and processing contention continue to degrade performance. A similar trend appears in the oversized-BSM flooding scenarios (i.e., valid 600-byte BSM flooding at 500 Hz and 1000 Hz) although the mitigation improves delivery performance, its relative benefit diminishes as the per-packet parsing and handling cost increases, because the host/application path remains loaded by more expensive message processing. Overall, Figure 9 suggests that lightweight host-side controls can delay the onset of collapse and improve robustness under moderate attack pressure, but they do not fully mitigate the underlying vulnerability when sustained arrival pressure and per-packet processing load become too large.

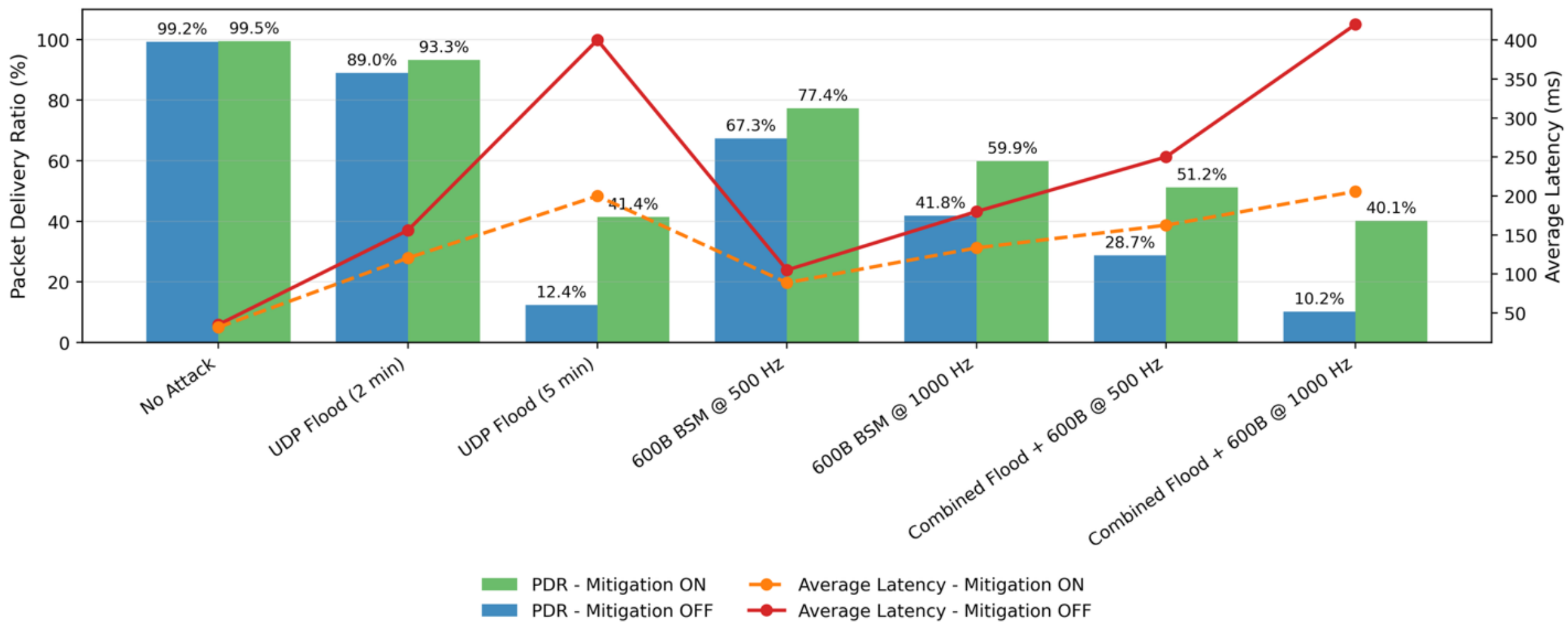

**Figure 9. Measured Performance with Token-Bucket + FQ-CoDel Mitigation OFF and ON**

To further evaluate the robustness of the implemented token-bucket + FQ-CoDel mitigation beyond the fixed attack cases, we next examine how performance evolves as the attack frequency increases. This sensitivity analysis is important because it reveals not only whether the mitigation improves average-case performance, but also how far it extends the operational boundary of the FCW communication pipeline before saturation occurs.

Figures 10(a) and 10(b) present the measured sensitivity of PDR and average latency, respectively, to increasing attack frequency with mitigation disabled and enabled. Without mitigation, performance degrades rapidly as attack frequency increases, with PDR falling sharply and latency rising well beyond the values required for timely FCW operation (i.e., latency of 100ms and PDR > 90%). With token-bucket + FQ-CoDel enabled, this degradation becomes more gradual: the receiver maintains consistently higher delivery performance and lower latency across the full range of tested attack rates, although the benefit decreases as the attack becomes more aggressive. This pattern is consistent with the role of host-side ingress control. Under moderate overload, early rate limiting and queue management help prevent deep backlog buildup and preserve receiver processing capacity. Under stronger sustained flooding, however, the remaining arrival pressure and per-packet processing burden still exceed the available host/application capacity, so performance continues to degrade even with mitigation enabled. These communication-level trends are also consistent with the FCW behavior observed in the unmitigated results (i.e., Table 2), where lower PDR and higher latency are associated with earlier last-valid-BSM breakdown and delayed or missed FCW triggering.

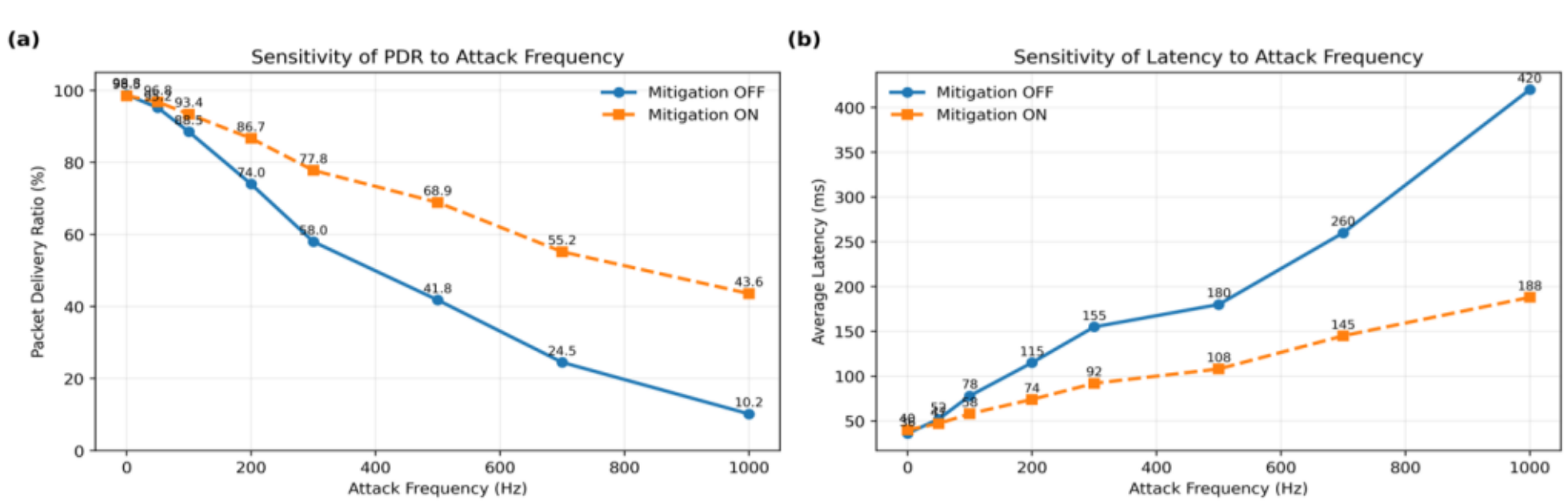

**Figure 10. Sensitivity Analysis of the Implemented Token-Bucket + FQ-CoDel Mitigation Under Increasing Attack Frequency (a) sensitivity of PDR to Attack Frequency, (b) Sensitivity of Latency to Attack Frequency**

Since any mitigation for C-V2X safety applications must be deployable on a resource-constrained OBU, we next quantify the implementation cost of the token-bucket + FQ-CoDel defense in terms of CPU load, per-packet processing time, defense-logic overhead, and memory usage.

Figures 11(a)–(d) summarize the computational overhead of the implemented token-bucket + FQ-CoDel mitigation in terms of CPU utilization, per-packet processing time, defense-logic overhead, and memory footprint, respectively. As shown in Figure 11(c) and Figure 11(d), the added defense logic remains bounded across all tested scenarios, with only a small per-packet overhead and a modest additional memory footprint. Under benign conditions, the mitigation introduces some expected computational cost, as reflected in the modest increases in CPU utilization and per-packet processing time shown in Figures 11(a**)** and 11(b). This behavior is reasonable because packets still traverse the policing and queue-management path even when no severe backlog is present.

As attack pressure increases, however, the interpretation becomes more nuanced. Under lighter attack conditions, the mitigation can still increase per-packet processing time because the added control logic is active while the receiver has not yet entered deep overload. Under more severe attack conditions, by contrast, the mitigation reduces the amount of abusive traffic that propagates deeper into the host/application pipeline, which lowers downstream parsing contention and limits persistent queue buildup. This effect is reflected in Figures 11(a) and **11(b)**, where CPU utilization and effective processing time decrease relative to the unmitigated cases under the strongest sustained and combined flooding scenarios. Overall, Figure 11 indicates that token-bucket + FQ-CoDel imposes a small and bounded overhead during benign operation, while under severe attack, it can partially offset overload-induced computation by discarding abusive traffic earlier in the processing path.

When combined, the results of Figures 9–11 show that the evaluated token-bucket + FQ-CoDel mitigation provides only partial protection against protocol-compliant flooding attacks. Although it improves PDR, reduces latency, and imposes only modest overhead under normal conditions, it does not fully prevent severe degradation under the strongest sustained and combined attack scenarios. A likely reason is that ingress policing and queue management can reduce abusive arrivals and backlog growth but cannot fully eliminate the remaining arrival pressure and per-packet processing burden once the receiver host/application path approaches saturation. These findings therefore reinforce the main conclusion of the paper: the commercial OBU-based FCW system evaluated in this study remains vulnerable to protocol-compliant flooding, even when lightweight host-side mitigations are applied. Future work should investigate more adaptive and application-aware defenses, such as tighter integration between ingress control and FCW scheduling, more selective packet prioritization, and multi-stage detection and mitigation mechanisms that can better preserve timely safety-message processing under sustained overload.

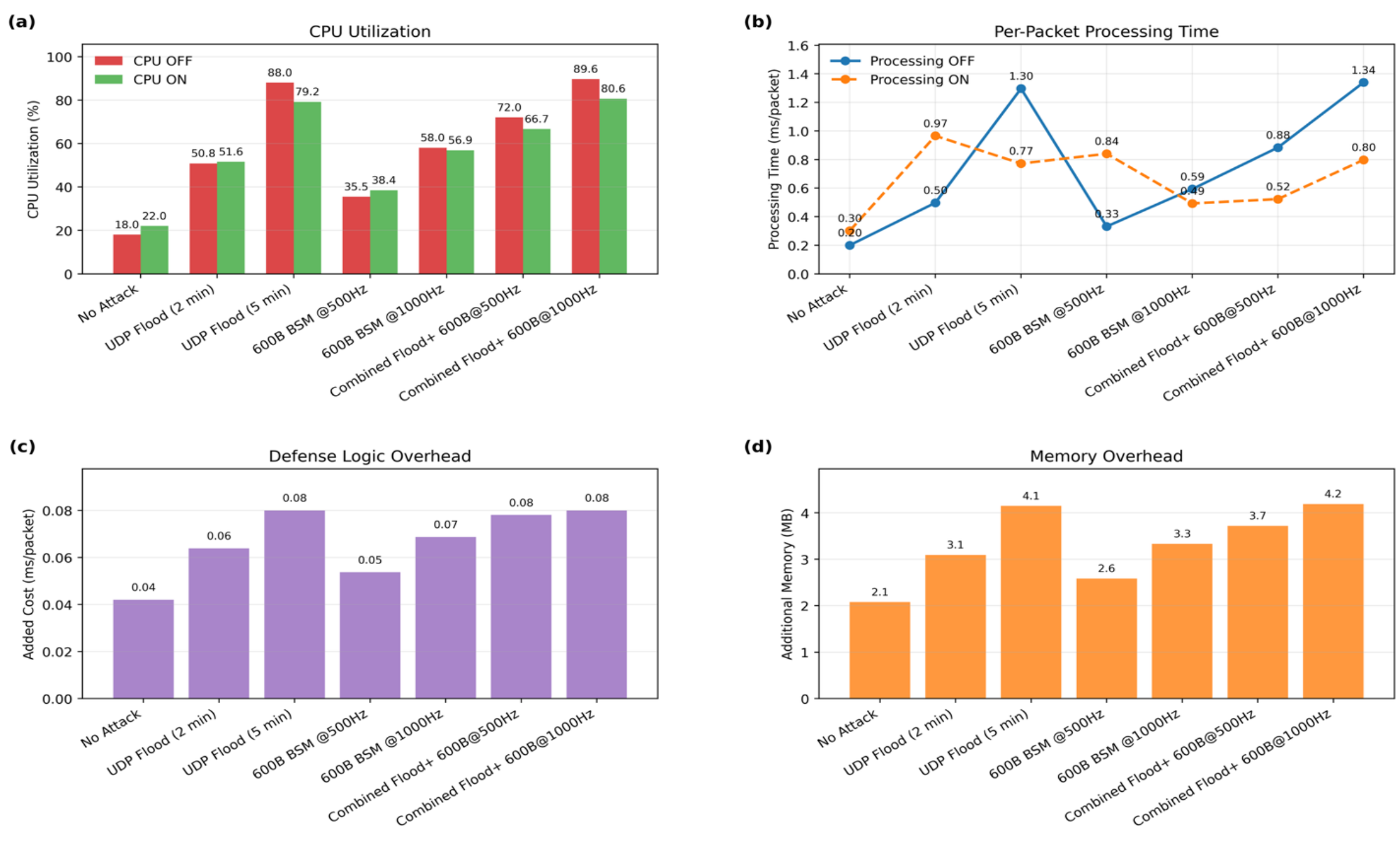

**Figure 11. Computational Overhead of the Implemented Token-Bucket + FQ-CoDel Mitigation (a) CPU Utilization, (b) Per-Packet Processing Time, (c) Defense Logic Overhead, (d) Memory Overhead**

## CONCLUSIONS AND FUTURE WORK

This study demonstrates that protocol-compliant flooding attacks pose a serious threat to C-V2X safety applications. Through real-world testing, we show that high-rate UDP floods and oversized valid BSM-like floods, individually and in combination, can push communication performance beyond safe operating thresholds and lead to delayed or entirely suppressed FCW alerts. Importantly, the dominant degradation observed in this testbed is governed mainly by the receiver host/application path, where high-rate valid traffic increases queue buildup, socket backlog, CPU contention, and FCW processing delay, rather than requiring malformed traffic or intrinsic saturation of the PC5 radio interface.

The results indicate that current security mechanisms, including those defined in the IEEE 1609 family security protocols, primarily address authentication, integrity, and RF-layer protections. They do not mitigate resource exhaustion caused by valid but overwhelming traffic (*41*). This exposes a gap in protocol and system-level defenses where authenticated, standards-compliant messages can still disrupt time-critical safety functions. Addressing this issue will require resilience improvements at the transport and the application layers, complementing existing IEEE 1609 protections.

Future work should explore multi-vector attacks that combine spoofing, jamming, and protocol-compliant flooding to provide a broader security assessment. Developing intrusion detection and prevention mechanisms that extend the IEEE 1609 security framework to handle resource exhaustion will be key. Large-scale field trials under varied traffic conditions and across different OBU platforms, as well as studies on other safety applications, such as cooperative braking and intersection collision avoidance, are necessary next steps. Ultimately, protocol enhancements and updated standardization efforts will be essential to ensure connected vehicle safety application remain reliable under stealthy flooding attacks in real-world deployments.

## ACKNOWLEDGMENTS

This work is based upon the work supported by the National Center for Transportation Cybersecurity and Resiliency (TraCR) (a U.S. Department of Transportation National University Transportation Center) headquartered at Clemson University, Clemson, South Carolina, USA. Any opinions, findings, conclusions, and recommendations expressed in this material are those of the author(s) and do not necessarily reflect the views of TraCR, and the U.S. Government assumes no liability for the contents or use thereof.

Moreover, the authors acknowledge using large language models (LLMs), particularly Google Gemini, to improve this manuscript. LLM was solely used for refining grammar, paraphrasing text and ensuring clarity of content.

## AUTHOR CONTRIBUTIONS

The authors confirm their contribution to the paper: J. Tine, M. Aldeen, A. Enan, M.S. Salek, L. Cheng, M. Chowdhury. All authors reviewed the results and approved the final version of the manuscript.

# Appendix A — Attacker-Side Algorithms

This section presents the algorithms that generate our attacks. Algorithm A-1 is the UDP flood generator and algorithm A-2 is the high-rate oversize BSM.

**Algorithm A-1**. UDP Flood Generator (Burst/Sustained, Minimal Payload)
Goal: Stress the victim's non-safety UDP ingress/queues while remaining protocol compliant .
Inputs: V_IP=192.168.93.100, V_PORT=9000, target rate PKT_HZ, duration D ∈ {120, 300} seconds.
Output: Minimal payload at PKT_HZ for D seconds.
As shown in Figure 12, this attack generates a steady, rate-controlled stream of small UDP packets to the receiver's non-safety ingress. Because the traffic is transport-layer protocol compliant (valid UDP/IP, no malformed frames), it does not violate radio or networking standards or alter the PC5 sidelink; instead, it increases socket backlog, copy/parse work and scheduler pressure in the host application path. By tuning PKT_HZ and pacing with interval = 1/PKT_HZ (with 120-s burst and 300-s sustained runs), we create a controlled load that causes latency growth according to rate of the attack. The 200-byte payload keeps packets lightweight, emphasizing per-packet overhead (queuing and dispatch) rather than raw bandwidth, which directly targets CPU/queue contention. The implementation is IPv4/IPv6-agnostic (auto-selecting AF_INET/AF_INET6 from V_IP), making the method topology-independent. In the context of our study, protocol compliance does not guarantee safety when edge resources are not prioritized. This design isolates application-layer resource exhaustion: even while genuine 10 Hz safety messages over PC5 continue uninterrupted, the challenging non-safety workload delays the FCW decision loop, increasing application latency and reducing the available time-to-warn.

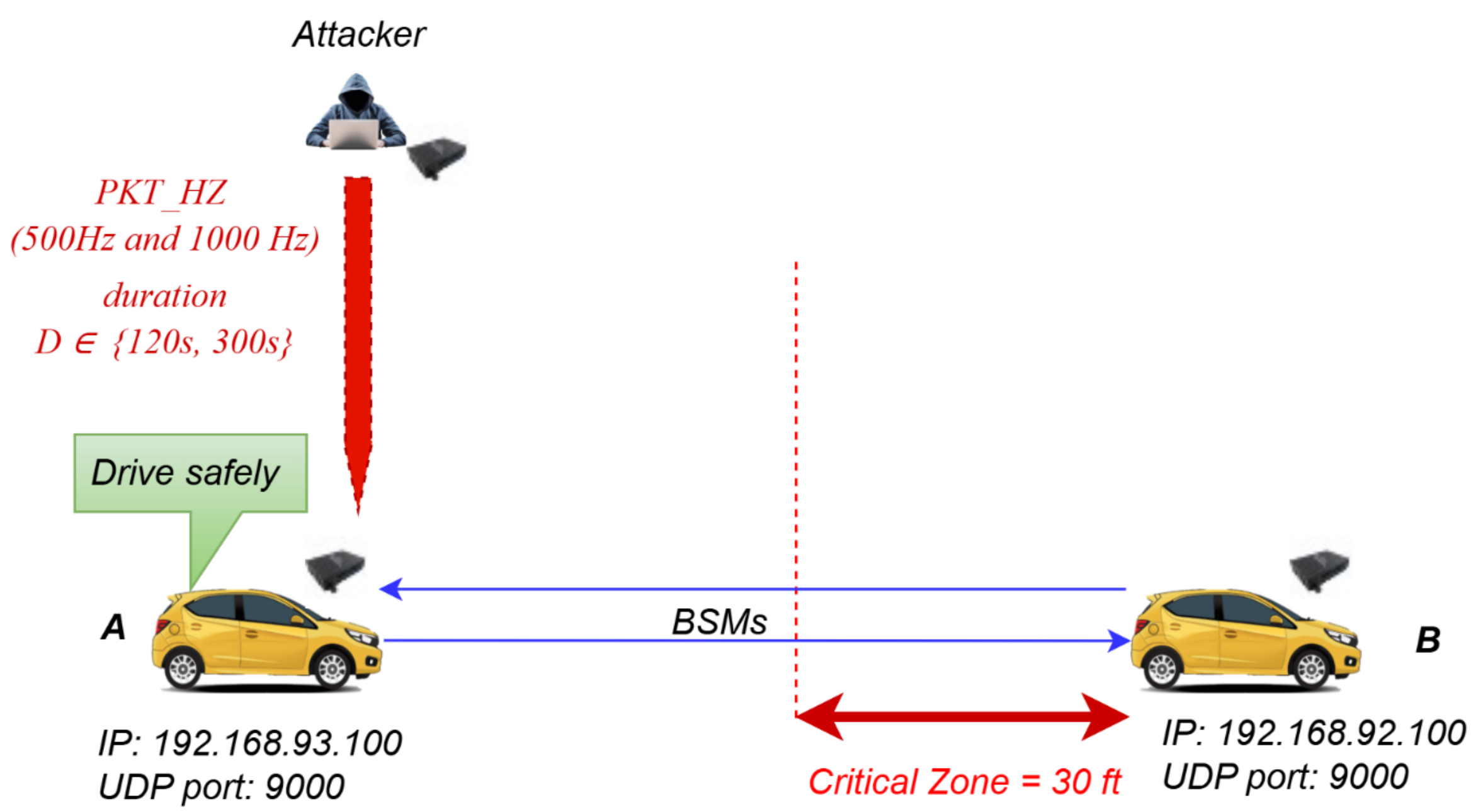

**Figure 12. UDP flood attack set up.**

**Algorithm 1 UDP Flood**

```
FAMILY ← AF_INET6 if ":" ∈ V_IP else AF_INET
SOCK socket (FAMILY, SOCK_DGRAM)
payload ← b"\x00"                // minimal payload (200 byte)
interval ← PKT_HZ               // we can change the frequency rate of the packet
t_end ← now () + D
while now() < t_end do
    sendto(SOCK, payload, (V_IP, V_PORT))
    sleep(interval)             // set to 0 for max − burst
end while
close (SOCK)
```

Notes: Use PKT_HZ high enough to reproduce the queue-growth/latency inflation as observed in Results; keep the transport layer compliant (i.e., no malformed frames).

**Algorithm A-2** — High-Rate Oversized BSM JSON Sender (500–1000 Hz)
Goal. Increase application-level parse cost and message arrival rate while remaining protocol compliant.
Inputs. V_IP = 192.168.93.100, V_PORT = 9000, RATE_HZ ∈ {500, 1000}, duration D (seconds), TARGET = 600 (bytes).
Output. Valid, minified JSON messages transmitted at RATE_HZ for D, each shaped like BSM content (including vehicle_id, latitude/longitude, speed, and timestamp) and padded with auxiliaries so that the encoded payload is equal to 600 bytes.
The generator constructs a minified JSON object with BSM-like fields, then computes padding to reach 600 bytes after encoding. High send rates (500/1000 Hz) increase parse/dispatch workload and contend for CPU/socket queues without violating transport standards.

**Algorithm 2 — High-Rate Oversized BSM JSON Sender**

```
FAMILY ← AF_INET6 if ":" ∈ V_IP else AF_INET
SOCK socket(FAMILY, SOCK_DGRAM)
interval ← RATE_HZ // the frequency rate
t_end ← now() + D; i ← 0
TARGET 600 // the payload size is set to 600 bytes
function ENCODE_MINIFIED(obj):
    // JSON without spaces to stabilize size
    return JSON_ENCODE(obj, separators = (",", ":"))  // language − specific
function MAKE_BASE(i, hist_len):
    base ← {
      "vehicle_id": "AttackerCar",
      "latitude":  ROUND(LAT0 + (i mod 5) · 1e − 6, 6),  // 6 decimals
      "longitude": ROUND (LON0 + (i mod 7) · 1e − 6, 6),  // 6 decimals
      "speed":     27.8,
      "time":      FLOOR (now()) // integer seconds
    }
    if hist_len > 0 then
```

```
      base["hist"] ← [ {"t": FLOOR(now()) − k, "s": 28} for k = 0…(hist_len − 1) ]
   end if
   return base
function MAKE_MSG(i):
   hist_len ← 4                    // small parser load
   obj ← MAKE_BASE(i, hist_len)
   // Compute remaining bytes for pad_str
   tmp ← ENCODE_MINIFIED(obj)
   // account for adding "pad_str": "" (key + quotes + colon + empty string)
   overhead
            ← LENGTH( ENCODE_MINIFIED({"pad_str": ""}) )
            + 2 // +2 for comma/braces
   remain ← TARGET − (LENGTH(tmp) + overhead )
   if remain < 0 then
     // too big already − reduce hist_len and recompute once
     hist_len ← MAX(0, hist_len − 2)
     obj ← MAKE_BASE (i, hist_len)
     tmp ← ENCODE_MINIFIED(obj)
     remain ← TARGET − ( LENGTH(tmp) + overhead)
   end if
   obj["pad_str"] ← REPEAT("X", MAX(0, remain))     // top up to target
   payload ← ENCODE_MINIFIED(obj)
   // Final nudge if off by a few bytes due to braces/commas
   diff ← TARGET − LENGTH(payload)
   if diff > 0 then obj["pad_str"] ← obj["pad_str"] + REPEAT("X", diff)
   if diff < 0 then obj["pad_str"]
            ← SLICE(obj["pad_str"], 0, LENGTH(obj["pad_str"]) + diff)
   return ENCODE_MINIFIED(obj)
while now() < t_end do
   msg ← MAKE_MSG(i)
   sendto(SOCK, msg, (V_IP, V_PORT))
   i ← i + 1
   sleep(interval)
end while
close(SOCK)
```

Notes: Use the same UDP access (port 9000) as the victim's FCW app. The JSON is valid and well-formed; the increase in rate and payload size stresses the receiver's parsing/dispatch loop without violating protocol.